\documentstyle[12pt,epsf]{article}
\newcommand{\r}{{\bf r}}
\newcommand{\x}{{\bf x}}
\newcommand{\k}{{\bf k}}
\newcommand{\rpe}{{\bf r_{\perp}}}
\newcommand{\rpl}{{\bf r_{\parallel}}}
\newcommand{\Beta}[2]{B\left(#1,#2\right)}
\newcommand{\vphi}{\vec{\phi}}
\def\beq{\begin{equation}}
\def\eeq{\end{equation}}
\def\beqa{\begin{eqnarray}}
\def\eeqa{\end{eqnarray}}

\begin{document}
\baselineskip=12pt
\begin{titlepage}
\noindent September 13, 1993 \\
\noindent Submitted to Phys.\ Rev.\ E \\
\noindent cond-mat/9310075
\vspace{3cm}
\baselineskip=24pt
\begin{center}
\begin{large}
{\bf PHASE ORDERING DYNAMICS OF THE O(n) MODEL - EXACT
PREDICTIONS AND NUMERICAL RESULTS}\\
\end{large}
\vspace{1.5cm}
{\em R. E. Blundell and A. J. Bray} \\
\vspace{1cm}
Department of Theoretical Physics \\ The University, Manchester M13 9PL \\
\vspace{1.5cm}
{\bf ABSTRACT }
\end{center}
\vspace{1cm}
We consider the pair correlation functions of both the order parameter field
and its square for phase ordering in the $O(n)$ model with nonconserved order
parameter, in spatial dimension $2\le d\le 3$ and spin dimension $1\le n\le d$.
We calculate, in the scaling limit, the exact short-distance singularities
of these correlation functions and compare these predictions to numerical
simulations. Our results suggest that the scaling hypothesis does not hold
for the $d=2$ $O(2)$ model.
\end{titlepage}

\newpage
\baselineskip=24pt
\pagestyle{plain}
\pagenumbering{arabic}

\section{Introduction}

When systems described by an order parameter with a continuous symmetry
are quenched from a disordered phase into a broken symmetry phase, the phase
ordering dynamics exhibited differs from that of systems with a discrete order
parameter in a number of important ways.
Many different systems in nature possess such symmetries, such as liquid
crystals, superfluids etc., and consequently there has been considerable
interest recently in adapting methods originally
developed to investigate discrete systems to encompass these new symmetries.

A major goal in the study of systems with both discrete and continuous
symmetries, which has been partially fulfilled in recent years, is the
approximate determination of the scaling function $f(x)$ for the two-point
correlation function. Although exact calculations have proved possible
only for physically uninteresting values of $d$ and $n$~\cite{RB4}, the spatial
and spin dimensionalities, approximate theories [2-7] have been proposed
which appear to agree closely with simulation results, both for the two-point
correlation function of the order parameter field $\vec{\phi}$,
\beq
C_2(\r,t)=\langle\vec{\phi}(\x,t)\vec{\phi}({\x+\r},t)\rangle=f(r/L(t))\ ,
\eeq
and its Fourier transform, the structure factor
\beq
S_2(\k,t)=L(t)^dg(kL(t)),
\eeq
where $L(t)$ is the characteristic length scale at time $t$ after the quench,
the angular brackets represent an average taken over initial conditions, and
the expected scaling forms have been anticipated in (1) and (2).
In addition, the two-point function of the square of the field,
\beq
C^U_4(\r,t)=\langle\vec{\phi}^2(\x,t)\vec{\phi}^2({\x+\r},t)\rangle
-\langle\vec{\phi}^2(\x,t)\rangle\langle\vec{\phi}^2({\x+\r},t)\rangle\ ,
\eeq
(where $U$ denotes `unnormalized') has recently been calculated within the
framework of these approximate theories~\cite{TopDefects}.
For scalar systems, the domain walls are sharp so
this correlation function is equivalent to the defect-defect correlation
function $\langle \rho(\x,t)\rho({\bf x+r},t)\rangle
- \langle \rho \rangle^2$, where $\rho(\x,t)$ is
the defect density at $\x$ at time $t$. Another case of physical interest is
that of a system with a complex scalar order parameter, e.g.\ superfluid
helium, which can be regarded as an $n=2$ vector system.
Here the correlation function of the order
parameter $\vec{\phi}(\r,t)$ is not experimentally accessible. Physical probes
couple to $\vec{\phi}^2(\r,t)$ instead and so $C^U_4(\r,t)$ is the
relevant correlation function to consider.

It is generally accepted that, for systems with a non-conserved order parameter
with $n\le d$, topological defects play an important role in the process of
phase ordering. For the $O(n)$ model, one finds that
the stable defects are interfaces for n=1, vortices for n=2 and monopoles for
n=3~\cite{Defects}. The idea we will use in the calculations presented in
section~\ref{exactSR} is that the order parameter field close to one of these
defects is determined solely by the geometric properties of the defect.  Bray
and Humayun have previously used this observation to calculate the singular
short distance behaviour of the two-point correlation function, which is
responsible for the power law tail in the structure factor, and are therefore
able to give an exact prediction of the {\em amplitude} of this tail~\cite{BH2}
as well as confirming the result $S(\k,t)\sim L^{-n}k^{-(d+n)}$
\cite{BrayPuri,Toyoki}.  The latter generalizes the familiar `Porod's Law'
\cite{Porod} to vector systems. We use
similar ideas to calculate the short distance form of $C_4(\r,t)$, the
correlation function of the square of the order parameter field, and in
section~\ref{comp} we compare simulation results to these predictions. We also
discuss the validity of the dynamic scaling hypothesis in relation for the
$d=2$
$O(2)$ model, both in relation to our short distance results and the simulation
data.

\section{Exact results}
\label{exactSR}
In this section we present exact calculations of the short distance behaviour
of the correlation functions discussed above. In principle, the idea behind the
calculations can be applied to most correlation functions, although there seems
little purpose in doing so as in general such correlation functions are not
experimentally accessible.

The calculation of the singular behaviour of the $C_2(\r,t)$ function at short
distance has, as we noted in the previous section, already been carried out by
Bray and Humayun, so we will only summarize the calculation here and refer the
reader to reference~\cite{BH2} for the details.

Consider the field $\vphi$ at points $\x$ and $\x+\r$ in the presence of a
point defect at the origin, appropriate to the situation $n=d$.  Let us denote
the defect density and average defect separation at time $t$ by $\rho(t)$ and
$L(t)$ respectively, and the defect core size by a.  Then for $a\ll |\x|\ll L$
and $a\ll |\x+\r| \ll L$ the field $\vphi$ at the two points will be saturated
in length and can be taken to be directed radially outward from the origin
\cite{Note1}. We therefore have
\beq
\vphi(\x)\cdot\vphi(\x+\r)=\frac{\x\cdot(\x+\r)}{|\x| |\x+\r|}.
\label{scalarprod}
\eeq
The singular short-distance behavior arises from the rapid spatial variation
in the direction of the vector $\vec{\phi}$ near the defect.
Holding ${\r}$ fixed and averaging over all possible positions of the defect
relative to the point $\x$ gives
\beq
C_{sing}(\r,t)=\rho(t)\int^L d^nx\,\left(\frac{\x\cdot(\x+\r)}{|\x||\x+\r|}
-{\rm analytic\ terms} \right).
\label{eq:a}
\eeq
The integral in equation~(\ref{eq:a}) is understood to be over all $\x$ which
satisfy $|\x|<L$. However, the subtraction of `analytic terms' from the
integrand (i.e. analytic in {\bf r}), sufficient to converge the integral
at large $|\x|$, enables us to evaluate the integral simply by extending it
over all space.

After some algebra the expression for $C_{\rm sing}$ obtained is
\beqa
C_{\rm sing}(\r,t)&=&nr^n\rho(t)\pi^{n/2-1}\Beta{\frac{n+1}{2}}{\frac{n+1}{2}}
\int_0^{\infty}du\,u^{-n/2-1}\,\left(e^{-u}-\mbox{analytic terms}\right)
\nonumber\label{eq:bm}\\
&=&nr^n\rho(t)\pi^{n/2-1}\Beta{\frac{n+1}{2}}{\frac{n+1}{2}}\Gamma(-n/2).
\label{eq:b}
\eeqa
Again, the `analytic terms' above refer to sufficient terms in the expansion
of $\exp(-u)$ in powers of $u$ to converge the integral. In equation (6),
$\Gamma(x)$ is the gamma function, and $B(x,y)$ the beta function.

For even $n$, the gamma function in equation~\ref{eq:b} contains poles.  To
deal with these cases, we set $n=2m+\epsilon$ and expand the resulting
expression, looking for the order unity term in $\epsilon$. We will discuss
this procedure more fully below in the context of the $C_4(\r,t)$ function.
Here we simply note that the principle effect is the introduction of an
additional factor of $\ln r$.

To generalize this result to $n \le d$ is straightforward. In this situation,
$\vphi(\r,t)$ lies in a plane normal to the defect and so we resolve $\r$ into
two parts; the component in this plane, $\rpe$, and the component parallel to
the defect core, $\rpl$.  The average over the orientations of $\rpe$ is
simply the calculation for the point defects above, and so the required result
comes from averaging equation~(\ref{eq:b}) over the orientations of $\rpl$,
with $\r$ replaced by its projection onto the plane, $\rpe$. The final result
for the singular short distance behaviour of the two-point correlation function
for general $n$ and $d$ is~\cite{BH2}
\beq
C_{\rm sing}(\r,t)=\pi^{n/2-1}\,
\frac{\Gamma(-n/2)\Gamma(d/2)\Gamma^2((n+1)/2)}{\Gamma((d+n)/2)\Gamma(n/2)}\,
\rho(t)r^n\ .
\label{eq:c}
\eeq
Once again, for even $n$ we expand the gamma function to extract the leading
singular term, which leads to a factor of $\ln r$. Taking the Fourier
transform of equation~(\ref{eq:c}) gives the large $k$ form of the structure
factor, valid for general $n \le d$,
\beq
S(\k,t)=\frac{\rho(t)}{\pi}(4\pi)^{(d+n)/2}\,
\frac{\Gamma^{2}((n+1)/2)\Gamma(d/2)}{\Gamma(n/2)}k^{-(d+n)}\ ,
\label{eq:d}
\eeq
valid for $ka \ll 1 \ll kL(t)$. We will compare this result to simulation
data later. Before we do so, let us consider the short distance
behaviour of the $\vphi^2$ correlation functions. Since $\vphi^2$ saturates
to unity at late times, however, it is in practice more convenient to
consider the pair correlation function of $1-\vphi^2$.
In particular, we are interested in this function normalized by its large
distance limit,
\beq
C_4(\r,t)=\frac{\langle(1-\vphi^2(\x,t))(1-\vphi^2(\x+\r,t))\rangle}
{\langle 1-\vphi^2(\x,t)\rangle\langle 1-\vphi^2(\x+\r,t)\rangle}.
\eeq
In soft spin models of Ising systems ($n=1$), the domain wall profile saturates
exponentially fast to $\pm 1$ and so the interfaces are sharp. This makes the
calculation of $C_4(\r,t)$ particularly simple. At any point in the system,
$\langle (1-\phi^2(\x,t))\rangle$ is zero unless it is close to a domain wall,
so we have immediately $\langle(1-\phi^2(\x,t))\rangle=a\rho(t)$ where this
relation defines $a$ as the domain wall width. For the product
$(1-\phi^2(\x,t))(1-\phi^2(\x+\r,t))$ to be non-zero requires
there to be an interface at $\x$ and at $\x+\r$.  If $r$ is small compared to
the characteristic distance between interfaces, $L(t)$, then the two points
must lie on the same interface. If we consider the point at $\x$ to lie on an
interface, then the additional requirement that $\x+\r$ lies on the same
interface gives a factor $a S_{d-1}/S_dr$, where $S_d$ is the surface area of
a $d$ dimensional unit sphere.  We have, therefore, for $n=1$,
\beq
\langle(1-\phi^2(\x,t))(1-\phi^2(\x+\r,t))\rangle=
a^2\frac{\rho(t)S_{d-1}}{S_dr}\ ,\ \ \ \ \ a \ll r \ll L(t)\ ,
\eeq
which gives, for $n=1$ and general $d$,
\beq
C_4(\r,t)=\frac{\Gamma(d/2)}{\pi^{1/2}\Gamma((d-1)/2)}\,\frac{1}{\rho(t)r}\ ,
\ \ \ \ \ a \ll r \ll L(t)\ .
\eeq
The $d=3$ result was recently derived by Onuki \cite{Onuki}.

For vector systems, the field far from a defect saturates as
$1-\vphi^2(\x)=a^2/x^2$, where $\x$ is the distance from the defect, and
$a$ is the `size' of the defect core. It follows, therefore, that the
short distance behaviour of $C_4(\r,t)$ is given by
\beqa
C_4(\r,t)&=&\frac{\langle(1-\vphi^2(\x,t))(1-\vphi^2(\x+\r,t))\rangle}
{\langle(1-\vphi^2(\x,t))\rangle\langle(1-\vphi^2(\x+\r,t))\rangle}
\nonumber\\
&=&\left(\rho\int^Ld\x\,\frac{a^4}{\x^2(\x+\r)^2}\right)
\Bigg /\left(\rho\int^Ld\x\,\frac{a^2}{\x^2}\right)^2\ ,
\label{c4:eq1}
\eeqa
provided the integrals are dominated by the field close to a defect.  As
before, we consider first the case of point defects, $n=d$, then
generalize the result to $n \le d$.  The denominator in
equation~(\ref{c4:eq1}) can be evaluated by changing to polar coordinates
and gives, for $n=2$,
\beq
C_4^D=(2\pi a^2\rho\ln(L/a))^2\ ,\ \ \ \ \ \ \ n=2\ .
\label{c4D}
\eeq
For $n>2$ the denominator in (\ref{c4:eq1}) is no longer dominated by the
field close to the defects and so we cannot treat these cases in the same
detail as for $n=2$. However, we can still evaluate the leading
short-distance behavior of the numerator for $2 <n \le 4$, and the leading
short-distance singularity of the numerator for any $n \ge 2$.

To evaluate the numerator (denoted by superscript `$N$'),
we first make a rescaling $\x\to r\x$ which gives
\beqa
C_4^N(\r,t)&=&\langle(1-\vphi^2(\x,t))(1-\vphi^2(\x+\r,t))\rangle
\nonumber\\
&=&\rho\,r^{n-4}\int_{a/r<x<L/r}d\x\,\frac{a^4}{|\x|^2|\x+\hat{\r}|^2}
\label{c4:eq2}
\eeqa
We now take $L/r\to\infty$ and $a/r\to 0$. For $n=2$, this makes the integrals
become logarithmically divergent, and the following treatment applies strictly
only for $2<n<4$, as we shall see below. We will discuss how the calculation is
extended to $n=2$ later.

Using the integral representation $\int_0^{\infty}dv\, \exp(-uv)=1/u$ gives
\beq
C_4^N(\r,t)=\rho a^4\,r^{n-4}
\int_0^{\infty}du\int_0^{\infty}dv\int_{-\infty}^{\infty}d\x\,
\exp(-u|\x|^2 - v|{\bf x+\hat{r}|}^2)\ .
\eeq
Evaluating the $\x$-integral first, followed by the $(u,v)$ integrals,
gives, after some algebra,
\beq
C^N_4(\r,t) = \pi^{n/2}\frac{\Gamma((4-n)/2)\Gamma^2((n-2)/2)}{\Gamma(n-2)}\,
\rho(t)a^4\,r^{n-4}.
\label{neqd}
\eeq

%
%
%
%

To extend this result to $n<d$ we consider the vector $\r$
to be resolved into two components, $\rpe$ and $\rpl$, where $\rpe$ is the
$(d-n)$-dimensional component of $\r$ in the plane perpendicular to the
defect core and $\rpl$ is the orthogonal vector containing the remaining
components of $\r$.  Using generalized spherical polar co-ordinates,
the projection of $\r$ into the plane perpendicular to the defect is simply
\beq
|\rpe|=r\prod_{i=1}^{d-n}|\sin\theta_i|
\eeq
To calculate $C^N_4(\r,t)$ we use the result from equation~(\ref{neqd})
with $\r$
replaced by $\rpe$ and average over all orientations of
$\r$, which is equivalent to averaging over the possible relative positions of
the two points to the defect. This gives
\beq
C^N_4(\r,t)=\pi^{n/2}\,\frac{\Gamma((4-n)/2)\Gamma^2((n-2)/2)}{\Gamma(n-2)}\,
\rho(t)a^4\,\langle |\rpe|^{n-4}\rangle.
\label{18}
\eeq
To evaluate this expression we require the average
\beqa
\left\langle|\rpe|^{n-4}\right\rangle&=&r^{n-4}\,
\left\langle\prod_{i=1}^{d-n}|\sin\theta_i|^{n-4}\right\rangle
\nonumber\\
&=&r^{n-4}\,\prod_{i=1}^{d-n}\left(
\int_0^{\pi/2}d\theta_i\,(\sin\theta_i)^{n-4+d-1-i}\right/\left.
\int_0^{\pi/2}d\theta_i\,(\sin\theta_i)^{d-1-i}\right)
\nonumber\\
&=&r^{n-4}\,\frac{\Gamma(n-2)\Gamma(d/2)}{\Gamma((n+d)/2-2)\Gamma(n/2)}.
\eeqa
Inserting this into (\ref{18}) gives, for general $d \ge n$
and $2<n<4$, the short-distance behavior
\beq
C^N_4(\r,t)=\pi^{n/2}\frac{\Gamma(2-n/2)\Gamma(d/2)
\Gamma^2(n/2-1)}{\Gamma((n+d)/2-2)\Gamma(n/2)}\,
\frac{\rho(t)a^4}{r^{4-n}}\ ,\ \ \ \ \ a \ll r \ll L(t)\ . \label{gen}
\eeq
In fact this result gives the leading short-distance singularity in
$C_4^N$ for all $n>2$ (with the usual logarithmic factors for even $n$,
which can be extracted by taking the limit from real $n$ appropriately).
For $2< n \leq 4$, it gives the leading short distance behavior
(not just the singularity), whereas, for $n>4$, $C_4$ saturates to a
constant at short distance. (By `short distance' we mean, of course,
$a \ll r \ll L(t)$ as usual). Equation (20) implies the power-law tail
$S_4(\k,t) \sim \rho(t)\,k^{-(d+n-4)}$, for $d \ge n >2$, in the Fourier
transform of $C_4$, a form first suggested in \cite{TopDefects}.

For $n=2$, there is a double pole in equation~(\ref{gen}) for $d\neq 2$ and a
single pole if $d=2$. These poles come from the lower end of the integration
range in equation~(\ref{c4:eq2}). In reality, $(1-\phi^2(\x,t))=a^2/x^2$ breaks
down at $x\approx a$ so the integrals in equation~(\ref{c4:eq2}) should be cut
off at $r \simeq a$. The effect of introducing the cutoffs would be to
eliminate the poles in equation~(\ref{gen}). To extend this result to $n=2$
then, we set
$n=2+\epsilon$ and expand the resulting expression to find the $O(1)$ term in
$\epsilon$. The terms which are divergent as $\epsilon\to 0$ would be removed
by a short distance cutoff. This gives, for $d > 2$,
\beqa
C^N_4(\r,t)&=&\rho(t)a^4\pi^{1+\epsilon/2}\frac{\Gamma(1-\epsilon/2)\Gamma(d/2)
\Gamma^2(\epsilon/2)}{\Gamma((\epsilon+d)/2-1)\Gamma(1+\epsilon/2)
r^{2-\epsilon}}
\nonumber\\
&\to&\frac{(d-2)\pi\rho(t)a^4\ln^2 (r/a)}{r^2},
\label{gend}
\eeqa
on retaining the $O(1)$ term in $\epsilon$. We introduce factors of the form
$\ln (r/a)$, rather than $\ln (r/L)$ say, because the cutoff required on the
integrals is at $r=a$.  For $d=2$, we need to take the limit $d\to 2$ before
taking $n\to 2$. This leads to a cancellation of two of the gamma functions in
equation~(\ref{gen}) leaving
\beqa
C^N_4(\r,t)&=&\rho(t)a^4\pi^{1+\epsilon/2}
\frac{\Gamma(1-\epsilon/2)\Gamma(\epsilon/2)}{\Gamma(1+\epsilon/2)
r^{2-\epsilon}}
\nonumber\\
&\to&\frac{2\pi\rho(t)a^4\ln (r/a)}{r^2}\label{c4n2}
\eeqa
on retaining the $O(1)$ term in $\epsilon$.  A direct calculation of the $n=2$
case with cutoffs on the integrals confirms these results, which are correct
to leading logarithmic accuracy.

In the scaling limit, where $r/L$ is kept fixed and $L\to\infty$,
equations~(\ref{gend}) and (\ref{c4n2}) can be simplified using
\mbox{$\ln(r/a)=\ln(r/L)+\ln(L/a)\approx\ln(L/a)$}. The normalized correlator
$C_4=C_4^N/C_4^D$ is therefore, using~(\ref{c4D}) for $C_4^D$,
\beqa
C_4(\r,t) &=& \frac{1}{2 \pi \rho \ln(L/a) r^2}\ ,\ \ \ \ \ d=2\ ,
\label{c4d2scaling} \\
          &=& \frac{d-2}{4\pi\rho r^2}\ ,\ \ \ \ \ d>2\ ,
\label{c4d3scaling}
\eeqa
valid for $a \ll r \ll L(t)$.
It is interesting that for $d>2$, the logarithms in $C_4^N$ and
$C_4^D$ cancel to produce an expression in a conventional scaling form
(with scaling variable $\rho r^2$), as would be expected,
whereas for $d=2$ a single logarithm remains in the
denominator and therefore $C_4(\r,t)$ does not have a scaling form at short
distance~\cite{TopDefects}.
This does not necessarily imply that there are no scaling phenomena
in the $d=2$ $O(2)$ model but it does show that at least this correlation
function does not scale. Taken together with evidence from the simulations
of ~\cite{ONSim}, and our own simulation results presented below, that
indicate a multiplicity of length scales in this system, as well
as the calculation of Bray and Rutenberg of the energy dissipation~\cite{BR},
it now seems likely that scaling does not occur in the $d=2$ $O(2)$ model.

In summary, we have exact predictions for the short distance singularity of
$C_2(\r,t)$, or equivalently, for the tail amplitude of the structure factor.
We have also predicted the short distance form of $C_4(\r,t)$ for $n\le 2$,
which for the 2d XY model violates the scaling hypothesis. We now compare
these results to computer simulations.

\section{Simulation results}

We now turn our attention to computer simulations of the $O(n)$ model, the
results of which we will compare to the above predictions.  For $n>d+1$ there
are no stable topological defects and for $n=d+1$ the defects are extended
textures, the dynamics of which are not well understood. We therefore have
restricted our investigations to  $2\le d \le 3$ and $1\le n \le d$. The data
from the $n=1$ simulations, some of which have already been published in
another form~\cite{BBS},  were provided by S. Sattler.

\subsection{Cell dynamical simulations}

The simulation of phase ordering using cell dynamical simulations
(CDS) is by now standard~\cite{CDS}.
The essential idea is to take a
lattice of `soft' spins, corresponding to a coarse-grained order
parameter field $\vphi(\r,t)$, and update them with the mapping
\beq
\vphi_{n+1}(i)=D \left( \frac{1}{z} \sum_j
  \vphi_{n}(j)-\vphi_{n}(i) \right ) +
E \hat{\vphi}_n(i) \tanh (|\vphi_n(i)|)
\label{CDS:eq}
\eeq
where $\hat{\phi}=\vphi / | \vphi |$ (for scalar fields
$\hat{\phi}= {\rm sgn} (\phi)$),
$z$ is the number of nearest neighbours and $D$ and $E$ are adjustable
parameters, for which we chose the standard values $D=0.5$ and
$E=1.3$~\cite{ONSim}. One can regard equation~(\ref{CDS:eq})
as a discrete TDGL equation in which the first term is the discretized
Laplacian and the second is derived from a potential $V(\vec{\phi})$, the
form of which has been chosen to keep the iteration process numerically stable
and to give the same attractors of the dynamics as the TDGL equation.

It is desirable when measuring $C_2$ to `harden' the
spins, which is achieved by computing (for general $n$)
\beq
C^H_2(\r,t)=\langle \hat{\phi}(\x,t) \cdot
\hat{\phi}({\bf x+r},t) \rangle .
\eeq
This reduces the effective size of the
defects (domain walls, strings etc.) to zero and hence scaling can be
achieved at smaller values of $r$.  Unfortunately,
the same procedure
cannot be applied to $C_4$ (which would vanish identically) and
hence the scaling in this case for small $r$ is considerably worse.

To measure the defect density in the vector systems we used two different
algorithms. For the $d=n$ systems, the defects are points and so can be
identified by looking for peaks in the local energy density $E_i=-\sum_{<j>}
\vphi_i \cdot \vphi_j$, where the sum is over the nearest neighbors of $i$.
For the $d=3$ $O(2)$ model, however, the
defects are strings and consequently this method is not so simple to use.
Instead, we look for plaquettes around which the order parameter
rotates through $\pm 2\pi$ radians and associate with each one a unit
length of string.  The same approach applied to the $d=2$ $O(2)$ model
gives results identical to those obtained with the algorithm based on
the energy density. It is important to note, however, that the length
of string given by this procedure is that given by a Manhattan metric,
because the string is constrained to lie on the dual lattice. To compare
to our calculations in section~\ref{exactSR}, which were based on a
continuum description, requires that we use a Euclidean metric to
measure the string length. Assuming that on a scale large compared to the
lattice spacing the strings are isotropically
oriented, we can convert between these two lengths by
adjusting by a factor $[L_{\rm M}/L_{\rm E}]=3/2$ for $d=3$, where the
brackets represent an average taken over all orientations of the string
and $L_{\rm M}$ and $L_{\rm E}$ are the Manhattan and Euclidean lengths
of the string.

The details of the simulations are shown in table~\ref{table1}. In all the
following plots, we will use the symbols circles, squares, triangles,
diamonds and crosses to represent progressively later measurement times. The
actual times to which these symbols refer can be read off from the
table.
\begin{table}
\begin{center}
\begin{tabular}{||c|c|c|c|c|c||} \hline
&&&&& \\
spatial dimension & 2 & 2 & 3 & 3 & 3 \\
spin dimension & 1 & 2 & 1 & 2 & 3 \\
&&&&& \\ \hline &&&&& \\
&&&&& \\
lattice size & 700 & 128 & 80 & 64 & 64 \\
realizations & 30 & 500 & 45 & 250 & 150 \\
&&&&& \\ \hline
&&&&& \\
measurement times & 400 & 80 & 100 & 40 & 40 \\
                  & 800 & 160 & 200 & 80 & 80 \\
                  & 1200 & 320 & 300 & 160 & 160 \\
                  & 1600 & 640 & 400 & 320 & 320 \\
                  & 2000 & 1280 & 500 & 640 & 640 \\
&&&&& \\ \hline
\end{tabular}
\end{center}
\caption{Details of lattice size and number and length of runs for the
numerical simulations.}
\label{table1}
\end{table}
\subsection{Dynamic scaling}
We start the analysis of the data by considering a test of the dynamic scaling
hypothesis. If this hypothesis is true, then the data for the two-point
correlation function $C_2(\r,t)$ can be collapsed onto a single curve for each
system by rescaling $r$ by the characteristic scale $L(t)$.  This scale can be
defined in a number of ways. For example, it could be defined as the value of
$r$ for which $C_2(\r,t)=1/2$,
or even as the value which gives the best overall
collapse of the data. A definition of this sort can be rather misleading,
however, because a free parameter has been introduced for each value of $t$. As
the form of $C_2(\r,t)$ is rather featureless, it is not too surprising that a
good collapse of the data can be obtained using methods such as these. An
alternative and superior method is to use some characteristic length scale that
is not directly related to $C_2(\r,t)$.  If the scaling hypothesis holds then
the defect density should scale as $\rho(t) \sim L^{-n}(t)$, and so it seems
logical to use $r \rho^{1/n}(t)$ as the scaling variable. For the $n=1$
systems we have used, instead of $\rho$, $\langle 1-\phi^2\rangle$ which
scales in the same way as the true defect density.
Figures~\ref{c2p-d2n1} to~\ref{c2p-d3n3} show the two-point correlation
function scaled in this manner along with the theoretical predictions of
OJK for the scalar systems and of Bray, Puri and Toyoki (BPT) for the
vector systems. The general form of these results is \cite{BrayPuri,Toyoki}
\beq
C_2(\r,t) = \frac{n\gamma}{2\pi}\,\left[B\left(\frac{n+1}{2},\frac{1}{2}\right)
\right]^2\,F\left(\frac{1}{2},\frac{1}{2};\frac{n+2}{2};\gamma^2\right)\ ,
\label{BPT}
\eeq
where $F(a,b;c;z)$ is the hypergeometric function, and $\gamma=\exp(-x^2/8)$,
with $x=r/L(t)$ the scaling variable. In the theoretical curves, the
horizontal scale has been adjusted to give the best fit.

Apart from $d=2$, $n=2$, the collapse of the data is
excellent. For $d=3$, $n=3$ there is some evidence for finite size
effects in the last data set, but otherwise the data scale well. We
conclude that our numerical results are entirely consistent with the
scaling hypothesis in these systems. It is clear, however, that for
the $d=2$ $O(2)$ model, i.e.\ the 2d~XY~model, the data do not scale.
Furthermore, there is no evidence that the data at later times are
approaching an asymptotic scaling curve. It is possible that the
system has yet to reach the scaling regime even at the latest times,
but considering how early the scaling regime is reached for the other
systems, it is also possible that the scaling hypothesis fails for
this system. It is interesting that the data can be collapsed to a
single curve by using values for $L(t)$ specially chosen to do so.
The quality of the
scaling achieved is as good as in any of the other systems (see
figure~\ref{c2l-d2n2}).
Bray and Rutenberg also find evidence that there may be scaling violations
in the 2d XY model~\cite{BR}. They conclude that energy dissipation due to the
ordering process occurs significantly on all scales between the vortex
core size and the inter-vortex spacing, suggesting there may not be a
single length scale $L(t)$ that characterizes the system morphology.
We recall that the exact short distance behaviour of $C_4(\r,t)$
calculated in the previous section is also not in a scaling form.
In all cases, the theoretical curves fit very well with the data. There are
small discrepancies, but considering the approximations made in the calculation
of these curves, it is rather surprising how close they appear to be to the
simulation data. We will see later, when we eliminate the free parameter
required to compare the theory to simulation results, that there are in reality
considerable discrepancies which are hidden by the introduction of the free
parameter.

\subsection{Growth laws}

The growth law for the characteristic scale in systems with a non-conserved
order parameter is expected to be $L(t)\sim t^{1/z}$, with
$z=2$~\cite{RB3,BR}.  By fitting a straight line to a log-log plot
of the defect density against time we find the values of $1/z$ shown in
table~\ref{GLz} which follow from the relation $L(t)\sim \rho(t)^{-n}$.
\begin{table}
\begin{center}
\begin{tabular}{||c|c|c|c|c|c||} \hline
&&&&& \\
spatial dimension & 2 & 2 & 3 & 3 & 3 \\
spin dimension & 1 & 2 & 1 & 2 & 3 \\
&&&&& \\ \hline &&&&& \\
$1/z$ & 0.46 & 0.37 & 0.46 & 0.44 & 0.45 \\
&&&&& \\ \hline
\end{tabular}
\end{center}
\caption{Growth laws for the characteristic scale in $O(n)$ systems. The values
were obtained from the decay of the defect density.}
\label{GLz}
\end{table}
After infinite time all the defects should have been removed from the system.
Consequently, we expect $\rho(t)$ plotted against $t^{-n/z}$ to be a straight
line passing through the origin.  Figure~\ref{GR0} shows these plots along
with a straight line fit.
To quote errors on the values of $1/z$ is somewhat misleading because the data
at different times come from the same system and are therefore highly
correlated.  In all cases, however, a fit with $z=2$ is unacceptable, even for
the $n=1$ systems! The reason for this is unclear. It could be that the
systems have yet to reach their asymptotic behaviour. A more likely
explanation  is the weak pinning effect on the defects, introduced by the
lattice, which becomes more important at later times as the driving forces
become weaker.  However, the data are consistent with zero defect density
for $t\to\infty$, so there is no direct evidence that defects become frozen
at late times, as they do for hard-spin vector simulations~\cite{BH4}.
It should be noted that previous simulations on vector systems also find
growth exponents consistently less then 1/2~\cite{ONSim,vecsims}.

For the 2d XY model, the growth laws for $\rho(t)^{-1/2}$
and $L(t)$, the length
scale required to collapse the $C_2(\r,t)$ function, are noticeably different.
The log-log plots shown in figure~\ref{growth} show clearly the different
power laws involved.
A least squares fit gives $\rho(t)^{-1/2}\sim t^{0.37}$
and $L(t)\sim t^{0.42}$. An
estimation of the errors on these exponents is difficult due to correlations
between data points. Mondello and Goldenfeld have previously simulated the
2d~XY model with rather longer run times than we used~\cite{ONSim}. Their data
also provides evidence for different growth laws for different length scales
for the time ranges they used.

\subsection{Short distance results}
\label{comp}
\subsubsection{The structure function tail amplitude}

We now consider the predictions of Bray and Humayun for the amplitude of the
power law tail in the structure factor. We present only our results for vector
systems as a comparison to Ising simulations has already been made~\cite{BH2}.
The scaling form for the structure factor is $S(\k,t)=L^d(t)g(kL(t))$ and we
expect a power law tail of the form $L(t)^{-n}k^{-(d+n)}$. For large $k$,
$S(\k,t)k^{d+n}/\rho(t)$ should therefore be the constant predicted by
equation~(\ref{eq:d}). Using $L(t)=\rho^{-1/n}(t)$, the scaling variable is
$k\rho^{-1/n}(t)$ which we plot against
$S(\k,t)k^{d+n}/\rho(t)=h(k\rho^{-1/n}(t))$, where $h(x)$ is a scaling
function.
Figures~\ref{s2p-d2n2} to~\ref{s2p-d3n3} show these plots along with the
analytic predictions for the large $k\rho^{-1/n}$ limit.
Note that we have plotted
$\langle S(\k,t)k^{d+n}/\rho(t)\rangle$ rather than
$\langle S(\k,t)k^{d+n}\rangle/\langle\rho(t)\rangle$ because the calculated
amplitude of the tail should be correct in each individual system.
In all three cases, the numerical data is consistent with the predicted values
of the tail amplitudes.  For the largest values of $k$, there is a tendency of
the data points to move up away from the predicted line. This is presumably
because the calculations were carried out using a continuous order parameter
field and at these large $k$, the effect of the lattice becomes noticeable.
The case $d=3$, $n=2$ shows a small but systematic discrepancy from the
predicted value at large $k$. Again we attribute this to a lattice effect.
In particular, the correction applied to the data to compare lengths
measured on the lattice (Manhattan metric) to those calculated on the
continuum (Euclidean metric) becomes inappropriate at lengths scales
comparable to the lattice spacing. This effect accounts for the sign of
the discrepancy in Figure~\ref{s2p-d3n2}.

An interesting feature of the simulation curves is that they always approach
their asymptotic constant value from above. For a scalar order parameter,
Tomita~\cite{Tomita} has accounted for this in terms of the curvature of
the interfaces, but we are not aware of any similar calculations for
vector fields. This same feature can also be seen when
the data is plotted in the more usual form $\ln(L^{-d}(t) S(k,t))$ against
$\ln(kL(t))$, although, because of the logarithmic scale, the effect is less
pronounced. For completeness, we present the data replotted in this form in
figures~\ref{s2p-d2n2-log} to~\ref{s2p-d3n3-log}, along with the theoretical
curve predicted by BPT.
The rescaling factor introduced into the theoretical curves in these plots was
the same as that used to give the best fit to the two-point correlation
function in real space. Because this factor was determined using the whole of
the two-point correlation function, the curves fit the data best at small $k$.
Note that, since the data approach their asymptotic straight lines from above,
there is a tendency in fitting data to assign too negative a slope to the
structure factor tail.

\subsubsection{$C_4$ for the scalar systems}

For $n=1$ and $n=2$ we have exact predictions for the short distance behaviour
of $C_4(\r,t)$, the correlation function of the square of the order parameter.
Plotting $C_4(\r,t)$ against $C_2(\r,t)$ gives a parameter-free scaling curve
that can be used to test theories of phase ordering~\cite{BBS}.  Here we will
consider the short distance behaviour of these functions.  From the previous
section, we have for $n=1$
\beqa
C_2(\r,t)&=&1-2\pi^{-1/2}\frac{\Gamma(d/2)}{\Gamma((d+1)/2)}\,\rho(t)r\ ,
\ \ \ \ \ a \ll r \ll L(t)\ ,
\nonumber \\
C_4(\r,t)&=&\frac{\Gamma(d/2)}{\pi^{1/2}\Gamma((d-1)/2)}\,\frac{1}{\rho(t)r}\ ,
\ \ \ \ \ a \ll r \ll L(t)\ .
\eeqa
and so
\beq
C_4^{-1}(\r,t)=\pi\,\frac{\Gamma((d-1)/2)\Gamma((d+1)/2)}
{2\Gamma^2(d/2)} (1-C_2(\r,t))
\label{c2c4-exact}
\eeq
in this regime. Therefore we expect a plot of $C^{-1}_4(\r,t)$
against $1-C_2(\r,t)$
to be linear near the origin with gradients $\pi^2/4$ and $2$ for $d=2$ and
$d=3$ respectively. Figures~\ref{c2c4-d2n1} and~\ref{c2c4-d3n1} show the
data plotted in this manner for the 2d and 3d $n=1$ systems, along with
a line of the appropriate gradient~\cite{note2}.
The data depart from the expected line for very small $r$ because we require
$a\ll r\ll L(t)$. It does appear that for asymptotically large times
the data will lie on the line we predict. It is also possible to calculate
$C_4^{-1}(\r,t)$ as a function of $1-C_2(\r,t)$ using the approximate
theories of OJK~\cite{OJK} and Mazenko~\cite{Mazenko} mentioned above.
{}From equation (\ref{BPT}) with $n=1$ one obtains
$C_2(\r,t)=(2/\pi)\sin^{-1}(\gamma)$ while \cite{TopDefects,BBS}
$C_4(\r,t)=1/\sqrt{1-\gamma^2}$, where $\gamma$ is the normalized
two-point function for a gaussian auxiliary field (see
references~\cite{TopDefects} and~\cite{BBS}).
Eliminating $\gamma$ gives the parameter-free relation
\beq
C_4^{-1}=\sin \left( \frac{\pi}{2}(1-C_2) \right).
\label{c2c4-OJK}
\eeq
This prediction is included in Figures~\ref{c2c4-d2n1} and~\ref{c2c4-d3n1}.
Plotting the data this way provides an absolute test of theory. Clearly,
the existing theories based on a gaussian auxiliary field are not
quantitatively accurate.
It is interesting to note, however, that for small $r$, i.e.\ small $1-C_2$,
equation~(\ref{c2c4-OJK}) gives the same linear relationship between $C_4^{-1}$
and $1-C_2$ as the exact small $r$ result in equation~(\ref{c2c4-exact}) does,
and, moreover, the coefficients are the same in the limit $d \to \infty$. It is
also clear from figures~\ref{c2c4-d2n1} and~\ref{c2c4-d3n1} that the data are
closer to the predicted form in equation~(\ref{c2c4-OJK}) for $d=3$ than for
$d=2$.  These facts add to the growing body of evidence that the OJK and
Mazenko theories are exact in the large $d$ limit~\cite{BH3}. Bray and
Humayun~\cite{BH3} have recently discussed a procedure by which the
leading order OJK result can in principle be systematically improved.
The present data will provide a good test of any improvements to the
theory.

\subsubsection{$C_4$ for the XY model}

{}From section~\ref{exactSR} we have, for n=2, the results given by
equations~(\ref{c4d2scaling}) and~(\ref{c4d3scaling}) for $C_4(\r,t)$ for $d=2$
and $d >2 $ respectively. Figure~\ref{c4:d3n2} shows the simulation data for
$C_4(\r,t)$ for the $d=3$ XY model plotted against the expected scaling
variable $\rho r^2$.  The data scale well for large $r$ and are consistent
with the predicted short distance behaviour, $C_4^{-1}(\r,t)=4\pi\rho r^2$.

We noted above that we cannot calculate $C_4(\r,t)$ for the $O(3)$ system
because $\langle 1-\vphi^2(\r,t)\rangle$ is not determined by the field
close to the defects. The result (\ref{gen}) for
$C_4^N(\r,t)=\langle(1-\vphi^2(\x,t))(1-\vphi^2(\x+\r,t))\rangle$
is still valid, however, although we cannot use it as an absolute
prediction because the defect core size $a$ is not known.
{}From equation~(\ref{gen}) we expect, for $n=3$,
$C_4 \sim 1/r$ for small $r$. The scaling function for $1/C_4$ should
therefore be
linear for small $r$, which is certainly consistent with the data shown in
figure~\ref{c4:d3n3} (recalling that the data break away from the true
scaling curve for small $r$, when the condition $r \gg a$ no longer holds).

For the $d=2$~XY~model, we predicted in section~\ref{exactSR}
that $C_4(\r,t)$ would
not scale for small $r$, but instead would have the form given by
equation~(\ref{c4d2scaling}). Figures~\ref{c4:d2n2-1} and~\ref{c4:d2n2-2}
show the data plotted in the form predicted, and against the naive scaling
variable $\rho r^2$.
In both cases the data appear to scale reasonably well. There is an ambiguity
in the definition of $L(t)$, the characteristic scale, used in
figure~\ref{c4:d2n2-1}, so we have chosen $L(t)=1/\sqrt{2\rho}$, simply because
this gives the best agreement with the prediction for the small $r$ behaviour.
Different choices of $L(t)$ still give $C_4^{-1} \propto r^2$ for small $r$,
but with different gradients.
Our simulation results are therefore consistent with the
predicted form of $C_4(\r,t)$, although they do not constitute an absolute
test.  It is interesting that, in contrast to the two-point function, the data
can be scaled using $\sqrt{\rho r^2}$.  The deviations from scaling predicted
by equation~(\ref{c4d2scaling}) are only logarithmic in $L(t)$, however,
which does not vary greatly for the range of $L$ available from the
simulation. The fact that $C_4(\r,t)$ can be approximately scaled using
the defect density, whereas $C_2(\r,t)$ cannot, may be because $C_4(\r,t)$
is more closely associated with the defects.
For $n=1$ systems, for instance, because the domain walls are
sharp, $C_4$ is essentially a defect-defect correlation function.

Just as for the scalar systems, we can plot $1-C_2(\r,t)$ against
$1/C_4(\r,t)$ to obtain a parameter-free test of the BPT
result~\cite{BrayPuri,Toyoki}, equation (\ref{BPT}), and the related
theory of references~\cite{LiuMazenko} and \cite{BH1}. The equivalent
result for $C_4$ was obtained in reference [8]: $C_4=F(1,1;n/2;\gamma^2)$.
Eliminating $\gamma$ between $C_4$ and $C_2$ gives $C_4^{-1}$ as a function
of $1-C_2$. Figures~\ref{newc2c4-d2n2} to~\ref{newc2c4-d3n3}
show the data plotted in this form [22].
The data do not fit at all well quantitatively to the theoretical
prediction, but do improve for larger $d$, as expected~\cite{BH3}.

\section{Conclusions}

We have performed numerical simulations of the O(n) model for $1 \leq n \leq d$
in dimensions $d=2$ and $d=3$.  Exact calculations for the short distance
behaviour of $C_2$ and $C_4$ have been compared to the simulation data and
satisfactory agreement found.  A particularly interesting feature of these
exact
results is that for the 2d XY model they predict a form for $C_4$ which cannot
be expressed in a scaling form, a fact already noted by Bray in the context
of his approximate calculation of $C_4$ \cite{TopDefects}.
Also, $C_2$ does not scale when plotted against
$r\rho(t)^{1/2}$.  We therefore suggest that there are logarithmic violations
of scaling in this model.

\bigskip
\bigskip

\noindent{\bf Acknowledgements}

\noindent We thank S. Sattler for providing the data for the $n=1$ systems,
and for discussions. REB thanks the SERC for financial support.

\newpage
\noindent {\bf Figure captions}
\begin{enumerate}

\item The two-point correlation function of the $d=2$ scalar model. The
data have been scaled using $\langle 1-\vphi^2 \rangle$ which is proportional
to the defect density (data courtesy of S. Sattler). The solid curve is the
OJK prediction,
with the horizontal axis scaled to give the best fit by eye to the data.
\label{c2p-d2n1}

\item The two-point correlation function of the $d=2$ $O(2)$ model. The
data have been scaled using the defect density.
\label{c2p-d2n2}

\item Same as Figure 1, but for the $d=3$ scalar model.
\label{c2p-d3n1}

\item The two-point correlation function of the $d=3$ $O(2)$ model. The
data have been scaled using the defect density. The solid curve is the
BPT prediction, scaled to give the best fit by eye.
\label{c2p-d3n2}

\item Same as Figure 4, but for the $d=3$ $O(3)$ model.
\label{c2p-d3n3}

\item The two-point correlation function for the $d=2$ $O(2)$ model. The
data have been scaled using the values L(t) which gave the best scaling.  The
solid curve is the BPT prediction, scaled to give the best fit by eye.
\label{c2l-d2n2}

\item Growth laws for the defect density in $O(n)$ models.
Moving left to right and top to bottom, $(d,n)$ takes the values
(2,1), (3,1), (2,2), (3,2) and (3,3).
\label{GR0}

\item Log-log plot of $\rho (t)^{-1/2}$ (top) and $L(t)$ (bottom) against time
for the $d=2$ $O(2)$ model.
The straight lines have gradients 0.37 and 0.42 (top to bottom).
\label{growth}

\item Structure function of the $d=2$ $O(2)$ model, with the power-law
tail scaled out. The line is the exact result of Bray and Humayun for
the amplitude of tail.
\label{s2p-d2n2}

\item Same as Figure 9, but with $d=3$, $n=2$.
\label{s2p-d3n2}

\item Same as Figure 9, but with $d=3$, $n=3$.
\label{s2p-d3n3}

\item Structure function of the $d=2$ $O(2)$
model. The line is the curve calculated by BPT, scaled to give the best fit
by eye.
\label{s2p-d2n2-log}

\item Same as Figure 12, but with $d=3$, $n=2$.
\label{s2p-d3n2-log}

\item Same as Figure 12, but with $d=3$, $n=3$.
\label{s2p-d3n3-log}

\item Plot relating the correlation functions $C_2(\r,t)$ and $C_4(\r,t)$ for
the 2d scalar model (data courtesy of S. Sattler).  The straight line is the
exact small $r$ prediction and the curve is the `OJK' prediction
\cite{TopDefects,BBS}.
\label{c2c4-d2n1}

\item Same as Figure 15, but for the $d=3$ scalar model.
\label{c2c4-d3n1}

\item The correlation function $C_4(\r,t)$ for the 3d~XY model. The
straight line is the predicted small $r$ behaviour.
\label{c4:d3n2}

\item The correlation function $C_4(\r,t)$ for the $d=3$ $O(3)$ model. The
small $r$ behaviour is expected to be linear at asymptotically late times.
\label{c4:d3n3}

\item The correlation function $C_4(\r,t)$ for the 2d~XY model. The
straight line is the predicted small $r$ behaviour. We have used
$L(t)=k\sqrt{\rho}$ where the parameter $k$ was chosen to give the
required gradient.
\label{c4:d2n2-1}

\item The correlation function $C_4(\r,t)$ for the 2d~XY model plotted
in the naive scaling form $1/C_4(\r,t)$ against $\rho r^2$.
\label{c4:d2n2-2}

\item Plot relating the correlation functions $C_2(\r,t)$ and
$C_4(\r,t)$ for the 2d XY model. The curve is the `BPT' prediction
\cite{TopDefects,BBS}.
\label{newc2c4-d2n2}

\item Same as Figure 21, but with $d=3$, $n=2$.
\label{newc2c4-d3n2}

\item Same as Figure 21, but with $d=3$, $n=3$.
\label{newc2c4-d3n3}

\end{enumerate}


\begin{thebibliography}{00}
\bibitem{note2} The same data were presented in a different way in \cite{BBS}.
The present method has the advantage that all the data appear on the plot,
and the interesting short-distance region is concentrated near the origin.
\end{thebibliography}
\end{document}